\documentclass[epj]{svjour}
\usepackage{amsmath,amssymb,amsfonts,graphicx,cite,algorithmic,algorithm,multirow}
\graphicspath{{graphics/}}
\makeatletter
\ifx\input@path\@undefined
\def\input@path{{graphics/}}
\else
\g@addto@macro\input@path{{graphics/}}
\fi
\makeatother
\newcommand{\program}[1]{\textsf{#1}}

\preprint{DESY 11-147\\MCnet-11-20}

\title{ExSample -- A Library for Sampling Sudakov-Type Distributions}

\author{Simon Pl\"atzer}

\institute{DESY, Notkestrasse 85, D-22607 Hamburg, Germany}

\date{\today}

\abstract{Sudakov-type distributions are at the heart of generating
  radiation in parton showers as well as contemporary NLO matching
  algorithms along the lines of the POWHEG algorithm.  In this paper,
  the C++ library \program{ExSample} is introduced, which implements
  adaptive sampling of Sudakov-type distributions for splitting
  kernels which are in general only known numerically. Besides the
  evolution variable, the splitting kernels can depend on an arbitrary
  number of other degrees of freedom to be sampled, and any number of
  further parameters which are fixed on an event-by-event basis.
   }

\begin{document}

\maketitle


\section{Introduction}

Parton shower Monte Carlo simulations as implemented in
\cite{Bahr:2008pv,Sjostrand:2007gs,Gleisberg:2008ta}, just to name few
of the recently developed codes, require a way to draw random variates
from a probability density
\begin{multline}
\label{eqs:sudakov-general}
\frac{{\rm d}S_P(\mu,q|Q;z;\xi)}{{\rm d}q\ {\rm d}^nz} =
\Delta_P(\mu|Q;\xi)\delta(q-\mu)
+\\ \theta(Q-q)\theta(q-\mu)P(q;z;\xi)\Delta_P(q|Q;\xi)
\end{multline}
when evolving from a hard scale $Q$ to a soft scale $q$ in the
presence of an infrared cutoff $\mu$, below which no radiation
occurs. Here, $\Delta_P(q|Q;\xi)$ is the Sudakov form factor,
\begin{equation}
\Delta_P(q|Q;\xi) = \exp\left(-\int_q^Q\int P(k;z;\xi) {\rm d}^nz\ {\rm d}k  \right)
\end{equation}
and $P(q;z;\xi)\ge 0$ is the splitting kernel describing the dynamics
of radiation at a scale $q$, along with $n$ other kinematic parameters
$z=(z_1,...,z_n)$ and in dependence on any further parameters
$\xi=(\xi_1,...,\xi_m)$. Examples of these parameters are momentum
fractions of incoming partons or invariant masses of the partonic
configuration from which the next emission is to be generated. The
most complicated information in terms of additional parameters is
certainly given by the full information on a phase space point of a
Born-type event from which real emission is to be generated in the
context of matrix element corrections
\cite{Seymour:1994df,Bengtsson:1986hr,Bengtsson:1986et,Norrbin:2000uu,Miu:1998ju}
or NLO matching using the POWHEG method which is originally described
in \cite{Nason:2004rx}.  We refer to the probability density defined
in eq.~\ref{eqs:sudakov-general} as the Sudakov-type distribution
associated to $P$.

Drawing random variates from ${\rm d}S_P$ by standard methods is in
general not feasible, as the integral entering the Sudakov form factor
would have to be evaluated numerically, and interpolated.  Though this
is indeed being done for example in the \program{FORTRAN} version of
\program{HERWIG} \cite{Corcella:2002jc}, this method ceases to be
applicable if the number of additional degrees of freedom or in
particular the number of additional parameters become large.

To this extent, current parton shower implementations reside on the
Sudakov veto algorithm which, {\it e.g.} has been discussed in
\cite{Sjostrand:2006za,Seymour:1994df,Buckley:2011ms,Platzer:2011dq}. The
Sudakov veto algorithm requires an overestimate $R$ to the splitting
kernel of interest $P$, $R(q;z;\xi)\ge P(q;z;\xi)$, and is defined by
\vspace*{3ex}
\hrule
\begin{algorithmic}
\STATE $Q_{\text{start}} \gets Q$
\LOOP
\STATE solve {\bf rnd}$=\Delta_R(q|Q_{\text{start}};\xi)\theta(q-\mu)$ for $q$
\STATE draw $z$ from $R(q;z;\xi)$
\IF{$q=\mu$}
\RETURN $(\mu,z)$
\ELSE
\RETURN $(q,z)$ with probability $P(q;z;\xi)/R(q;z;\xi)$
\ENDIF
\STATE $Q_{\text{start}}\gets q$
\ENDLOOP
\end{algorithmic}
\hrule
\vspace*{3ex} where {\bf rnd} denotes a source of random numbers
uniformly distributed on $[0,1]$. Obviously, $R$ needs to be of a
simple form in such a way that the first step in the loop can easily
be implemented.

Finding such an $R$ has up to now always required knowledge of
properties of the target kernel $P$, making a general-purpose
implementation of the algorithm impossible. Especially towards more
complicated splitting kernels, this manual procedure of determining
$R$ from the properties of $P$ may not be possible at all: even
analytic expressions may not be known, $P$ being available only
numerically. A general implementation may also further enhance
flexibility when changing parton distribution functions in the parton
shower backward evolution and thus the respective splitting kernels.

The purpose of \program{ExSample} (a shorthand for
\program{Ex}ponential \program{Sample}r) is to provide such a general
purpose implementation, by adaptively obtaining an overestimate to the
target splitting kernel in such a way as to optimize the algorithm's
overall performance.

\section{Generation of Adapting Overestimates}
\label{sections:adapting-overestimates}

\program{ExSample} is very much inspired by the \program{ACDC} and
\program{FOAM} algorithms implemented in
\cite{Lonnblad:ACDC,Jadach:2002kn}.  By the same reasoning,
\program{ExSample} makes use of `cells', which represent a
sub-hypercube of the volume spanned by the evolution variable $q$, the
additional degrees of freedom $z$ and external parameters $\xi$. Cells
are organized in a binary tree, each cell having either two or no
children, in the latter case terminating the tree at this branch. The
union of the two hypercubes $U_b$ and $U_c$ represented by the two
children cells $c_{b,c}$ always equals the hypercube $U_{(bc)}$
represented by the parent cell $c_{(bc)}$. Each cell $c$ contains the
maximum of the target splitting kernel $P$ encountered by a
presampling as its value $w_c$. The leaf cells of the tree,
constituting a certain fractal-type partition of the sampling volume
into hypercubes, define the overestimate function,
\begin{equation}
R(q;z;\xi) = \sum_{\text{leaf cells}\ c} w_c\ \theta\left((q;z;\xi)\in U_c\right) \ .
\end{equation}
Each parent cell keeps track of the integrals of its children cells,
$I_{c,b} = w_{c,b} \text{volume}(U_{b,c})$.  This allows for an
efficient sampling of the overestimate function, by selecting either
of the children cells according to their integral, biased by
constraints imposed due to the selected evolution variable, the
externally fixed parameter point and the need to compensate for newly
encountered maxima.

\begin{figure}
\hfill\includegraphics[scale=0.4]{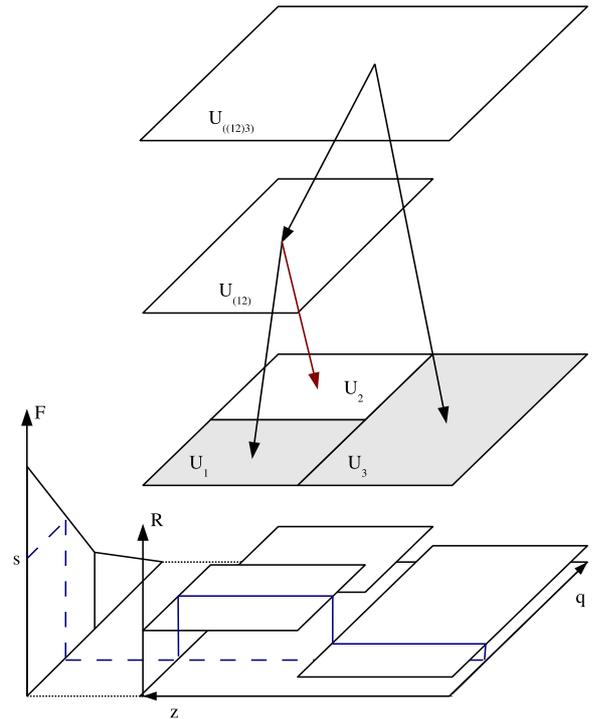}
\caption{\label{figures:trees-sudakov}A sketch of the algorithm for an
  evolution variable $q$, one additional variable $z$, and no further
  parameters $\xi$. The top of the figure shows how the leaf cells (in
  the third plane from the top, shown here after two cell splits) are
  organized in a binary tree structure starting from the root cell
  $U_{((12)3)}$. The bottom of the figure sketches the overestimate
  $R$. To the left of the overestimate, the Sudakov exponent
  corresponding to $R$, $F(q)=-\ln \Delta_R(q|1)$ is shown. Here we
  assume that the absolute upper bound on the evolution variable is
  $q<1$, thus the first step to draw an event starting from a scale
  $Q$ is to solve $s(Q)=-\ln {\bf rnd}+F(Q)=F(q)$ for q (indicated by
  the dashed blue line). A $z$ value is then sampled in the cells
  containing the $q$ value just chosen: The cell integrals over $z$
  are computed to only reflect the subtree consisting of the black
  arrows, and the tree structure is traversed only along the
  corresponding paths, selecting either of the children cells with
  weight given by the respective integral. Within the boundaries of a
  leaf cell selected by this procedure, a $z$ value is drawn flat.
  This corresponds to drawing a $z$ value from the distribution
  sketched by the solid blue line, the overestimate $R$ at fixed $q$.}
\end{figure}

The next value of the evolution variable is easily generated by
keeping track of projections of the overestimate kernel onto the
evolution variable dimension in dependence on the externally fixed
parameter point.  In order to keep track of the dependence on the
additional parameters $\xi$ as well as the starting value of the
evolution variable $Q$, \program{ExSample} provides a mechanism to
calculate unique hash values identifying the sub tree of the cell
structure which should be considered for a given parameter point. All
information needed to sample events, {\it i.e.}  in particular
projections of the overestimate kernel $R$ and the number of `missing'
events per cell, to be discussed in
section~\ref{sections:compensation}, can be accessed in dependence on
these hash values. The basic structure of the sampling is sketched in
figure~\ref{figures:trees-sudakov}.

The root cell of the tree spans the whole sampling volume and is the
only cell present at the initial stage of the algorithm. Children
cells are produced in an adaption step, iteratively building up the
cell tree through splitting a cell into two children cells. This
procedure aims at improving the algorithm's efficiency along with
gaining more detailed information on the target splitting kernel, {\it
  i.e.} a more fine-grained overestimate closer to it.

In order to achieve this, each cell always monitors its efficiency,
which is defined as the ratio of the number of accepted points divided
by the number of proposed points and thus gives a measure of the
overall performance of the Sudakov veto algorithm. If this efficiency
drops below a user-supplied threshold, the cell is considered
`bad'. With a frequency decreasing as the efficiency of the algorithm
increases, and on encounter of a bad cell, a potential splitting of
the cell is determined to further increase the efficiency of the
algorithm.

To obtain an optimal hyper-plane along which the cell should be split,
each cell histograms projections of the average target kernel value
onto each variable dimension $k$, $\langle P\rangle_k(x_k)$.  The
dimension $k$ (which here may refer to either the evolution variable,
one of the additional degrees of freedom or any of the external
parameters) orthogonal to this hyperplane, and the split point $x_k$
(out of a number of possible split points well inside the cell's boundaries)
defined by the intersection of the hyperplane and this direction are
determined to maximize a `gain' measure, defined as
\begin{equation}
\label{equations:splitgain}
\text{gain}_k(x_k) = \frac{\left|\int_{x_{k}^-}^{x_k}\langle P\rangle_k(x) {\rm d}x 
- \int_{x_k}^{x_{k}^+}\langle P\rangle_k(x) {\rm d}x \right|}{
\int_{x_{k}^-}^{x_{k}^+}\langle P\rangle_k(x) {\rm d}x
} \ .
\end{equation}
Here, $x_k^\pm$ denote the cell's boundaries in the variable
$x_k$. For reasons of performance and simplicity, the current
implementation uses a two-bin histogram per dimension, and $x_k =
(x_k^-+x_k^+)/2$, leaving only the choice of dimension to maximize the
gain measure: If the behaviour of $P$ is rather flat when projected on
one dimension $k$, this dimension will receive a small gain measure,
and projections showing more variance in $P$ are more likely to be
split along. Again, a user-supplied parameter can steer the behaviour
of the adaption by considering only those splits to be worth
performed, if the gain exceeds some value.

Out of the two children cells the target density is being presampled
in that cell which did not contain the maximum point used before to
get a new estimate of the maximum. The number of presampling points
per cell is another user-defined parameter. The choice of this
parameter has to be carried out in view of the compensation procedure
to be defined in the next section with a trade-off between the time
needed for presampling and the time lost by the number of events to be
vetoed by the compensation procedure. There is no general rule on how
it is to be determined. Experience gained so far shows that few
thousand presampling points are an acceptable compromise.

\section{Compensating for New Maxima}
\label{sections:compensation}

Since the true maximum of the target kernel can never be determined
with probability one from the presampling procedure, care has to be
taken on what constraints need to be imposed on the sampling procedure
once a point has been encountered exceeding the currently used
maximum.  For a sufficiently large number of presampling points one
may reside on the statement that these points are rare and generated
distributions will not show any effect on the erroneous
overestimate. Thinking about the overall efficiency of the algorithm
in performing its function of acting as a continuous source of
unweighted events with the smallest possible overhead, this is
certainly not a criterion to base an implementation on.

\begin{figure}
\hfill\includegraphics[scale=0.4]{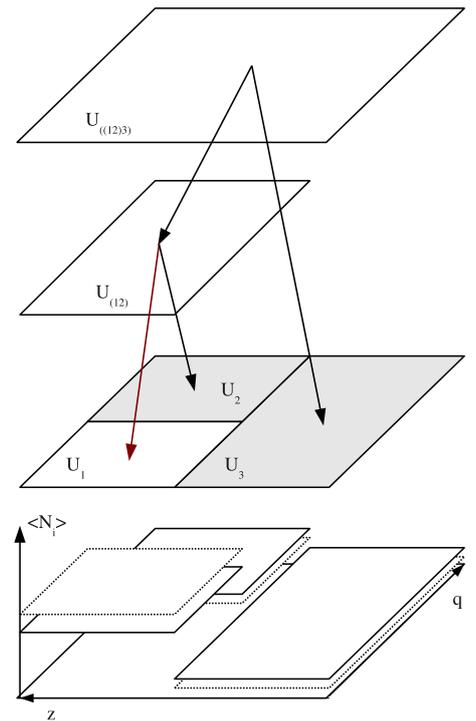}
\caption{\label{figures:trees-compensation}A sketch of the algorithm
  in a setup similar to figure~\ref{figures:trees-sudakov}, now
  sketching the situation upon encounter of a new overestimate. The
  new overestimate gave rise to different numbers of events expected
  in each cell (solid rectangles in the lower part), as compared to
  the number of events expected with the old overestimate (dotted
  rectangles). The difference between these determines the number of
  missing events per cell (see text for more details). In the sketch
  given here, the cells $U_2$ and $U_3$ would receive a positive number of
  missing events (forcing sampling in these cells as indicated by the
  black arrows), whereas cell $U_1$ would contain a negative count
  of missing events, triggering vetoes of attempts to sample points in
  this cell (indicated by the red arrow).}
\end{figure}

To define the method of compensation, we first introduce the notion of
missing events in a given cell. As for the cell's integral, each
parent cell carries the sum of the missing events of its children
cells. The number of missing events is not limited to be positive.  In
case it is positive, the corresponding cell needs to be oversampled,
{\it i.e.} the algorithm is forced to sample events in cells with a
positive number of missing events, lowering this number in the
selected cell if it is larger than zero. Oversampling is imposed on
the algorithm as long as there are cells with a positive count of
missing events. Conversely, if the missing event count is negative, a
cell needs to be undersampled. If such a cell is selected, its missing
event count is increased, if it is smaller than zero and the selection
is vetoed, triggering a new cell selection. The behaviour of the
algorithm in a compensating state is illustrated in
figure~\ref{figures:trees-compensation}.

Upon encounter of a new maximum $w_c' > w_c$, the number of missing
events associated to this change is calculated for each cell as
\begin{equation}
\label{equations:nmiss}
N^{\text{miss}}_c = N_c \left( \frac{p'_c}{p_c} - 1\right) \ .
\end{equation}
Here, $N_c$ is the number of proposal events already generated in the
cell, and $p_c$ ($p'_c$) denotes the probability to select cell $c$
using the old (new) overestimate value for events above the infrared
cutoff. $p_c$ is calculated from the knowledge of projections of the
overestimate kernel in dependence on the additional parameter point
$\xi$ and the hard scale $Q$ as
\begin{equation}
p_c = \frac{\int_{c} R(q;z;\xi) \Delta_R(q|Q;\xi){\rm d}^nz\ {\rm d}q}{1-\Delta_R(\mu|Q;\xi)} \ .
\end{equation}
$N^{\text{miss}}_c$ is then added to each cell's current missing event
count. Note that undersampling, $N^{\text{miss}}_c<0$ appears in the
cells not containing the newly encountered maximum owing to the change
in normalization of the overestimate density for events above the
infrared cutoff.  Eq.~\ref{equations:nmiss} ensures that within the
currently accumulated statistics proposal events are always
distributed according to the last encountered maximum, provided the
algorithm has been stopped in a state where it is not anymore forced
to perform over- or undersamplings.  This is evident by rewriting
eq.~\ref{equations:nmiss} as
\begin{equation}
N^{\text{miss}}_c = \frac{N_c}{\langle N\rangle_c} 
\left( \langle N'\rangle_c - \langle N\rangle_c \right)
\end{equation}
where $\langle N\rangle_c = N p_c$ ($\langle N'\rangle_c = N p'_c$) is
the number of expected events in cell $c$ for the total number of
generated events, $N$.  The difference in brackets is the number of
missing events in the absence of fluctuations due to a finite number
of generated events, and the factor in front of it takes into account
the currently accumulated statistics, {\it i.e.}  how much the
population of the cell differs from its expected population.

\section{The Cell Selection Algorithm}
\label{sections:cellselection}

In this section the complete algorithm to generate events as
implemented in \program{ExSample} is defined. We here skip those parts
connected to monitoring the efficiency of and splitting a cell.
Proposal events according to ${\rm d}S_R(q|Q;z;\xi)$ as required by
the Sudakov veto algorithm are generated by first deciding, if there
has been an event at the infrared cutoff or otherwise selecting a
proposal cell according to algorithm~\ref{algorithms:selection}.

\begin{algorithm}
\begin{algorithmic}
\STATE calculate sub tree hash $h(Q;\xi)$ and collect projections
\LOOP
\STATE solve {\bf rnd}$=\Delta_R(q|Q;\xi)\theta(q-\mu)$ for $q$
\IF{$q=\mu$}
\RETURN event at infrared cutoff
\ENDIF
\STATE collect cell integrals and missing event counters
\STATE cell $\gets$ root cell
\WHILE{cell is not a leaf}
\IF{$N^{\text{miss}}_{\text{firstChild(cell)}} > 0 \wedge N^{\text{miss}}_{\text{secondChild(cell)}} \le 0$}
\STATE cell $\gets$ firstChild(cell)
\ELSIF{$N^{\text{miss}}_{\text{firstChild(cell)}} \le 0 \wedge N^{\text{miss}}_{\text{secondChild(cell)}} > 0$}
\STATE cell $\gets$ secondChild(cell)
\ELSE
\IF{$\mathbf{rnd} < I_{\text{firstChild(cell)}}/I_{\text{cell}}$}
\STATE cell $\gets$ firstChild(cell)
\ELSE
\STATE cell $\gets$ secondChild(cell)
\ENDIF
\ENDIF
\ENDWHILE
\IF{$N^{\text{miss}}_{\text{cell}} = 0$}
\RETURN cell
\ELSIF{$N^{\text{miss}}_{\text{cell}} > 0$}
\STATE $N^{\text{miss}}_{\text{cell}} \gets N^{\text{miss}}_{\text{cell}} - 1$
\RETURN cell
\ELSIF{$N^{\text{miss}}_{\text{cell}} < 0$}
\STATE $N^{\text{miss}}_{\text{cell}} \gets N^{\text{miss}}_{\text{cell}} + 1$
\ENDIF
\ENDLOOP
\end{algorithmic}
\caption{\label{algorithms:selection} The cell selection algorithm.}
\end{algorithm}

Once a proposal cell has been selected, a proposal event is drawn by
sampling the remaining degrees of freedom $z$ in the selected cell
with uniform distribution.  Except for the compensating cell selection
algorithm outlined above, the Sudakov veto algorithm proceeds without
modification.

\section{Examples and Validation}
\label{sections:examples}

\program{ExSample} has been validated for various `toy' splitting
kernels and within the realistic application of a parton shower and
POWHEG matching implementation. In this section we present simple
examples of distributions obtained by using \program{ExSample}, mainly
to illustrate the basic functionality.

\begin{figure}
\centering
\input{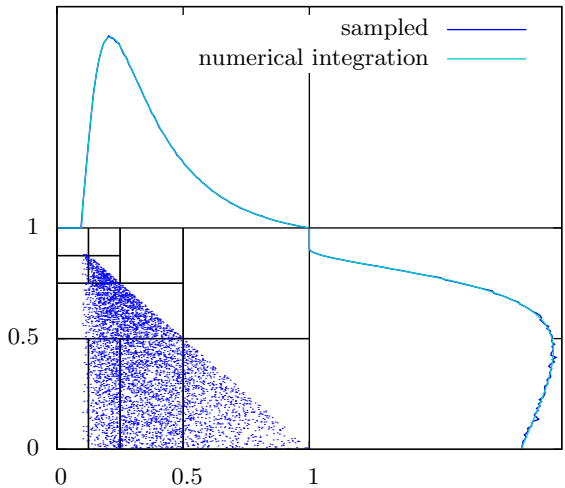}
\caption{\label{figures:samplesudakov}A Sudakov-type distribution with
  a QCD splitting function type kernel density as sampled by
  \program{ExSample} using the adaptive Sudakov veto algorithm. The
  vertical axis corresponds to the evolution variable $q$, the
  horizontal to a variable similar to a momentum fraction.  Shown are
  few sampled events, projections of the generated distribution versus
  the result from a numerical integration, and the the cell grid
  produced.}
\end{figure}

\begin{figure}
\centering
\input{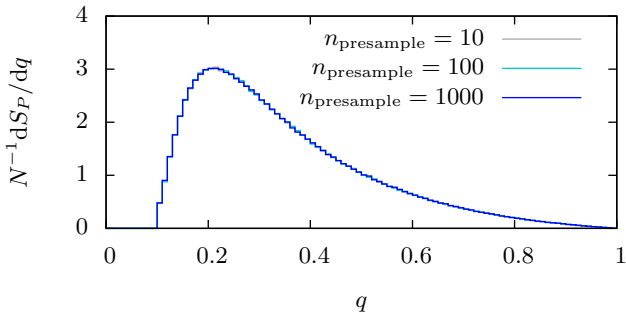}
\caption{\label{figures:compensatesudakov}The same distribution as
  shown in the upper left panel of figure~\ref{figures:samplesudakov},
  now sampled with a different number of presampling points proving
  functionality of the compensation procedure.}
\end{figure}

Figure~\ref{figures:samplesudakov} shows the results obtained by the
adaptive Sudakov veto algorithm, using a kernel density showing the
generic behaviour of a QCD splitting function with running $\alpha_s$.
Perfect agreement with a numerical integration is found. In addition,
figure~\ref{figures:compensatesudakov} shows the functionality of the
compensation procedure by comparing results for the same distribution
but different numbers of presampling points used in the algorithm,
which are all consistent with each other.

\begin{figure}
\centering
\input{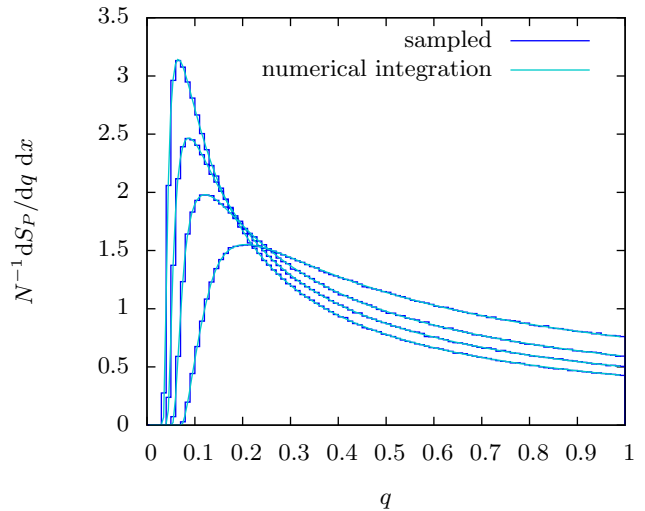}
\input{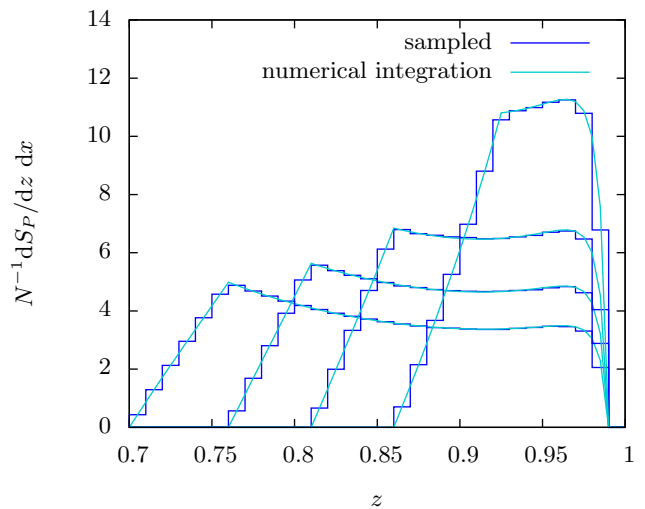}
\caption{\label{figures:sudakovx}Distributions for a Sudakov-type
  distribution using a quark splitting function, multiplied by a power
  law in $x/z$. Shown are the sampled distribution for the evolution
  variable $q$ and momentum fraction $z$ in various bins of the
  additional parameter $x$. The distributions are compared to a
  numerical integration, proving full functionality of the sampling in
  presence of additional parameters.}
\end{figure}

In figure~\ref{figures:sudakovx} the results of sampling a
Sudakov-type distribution in the presence of additional parameters are
shown. In this example, a quark splitting function multiplied by a
power law in $x/z$ has been used, where $x$ is the additional
parameter and $z$ is the momentum fraction variable to be sampled. The
sampled distributions in various bins of the additional parameter $x$
have been compared to a numerical integration. Full agreement has been
found here. The presence of adaption splits in the parameter dimension
has explicitly been checked for this example.

\begin{figure}
\centering
\input{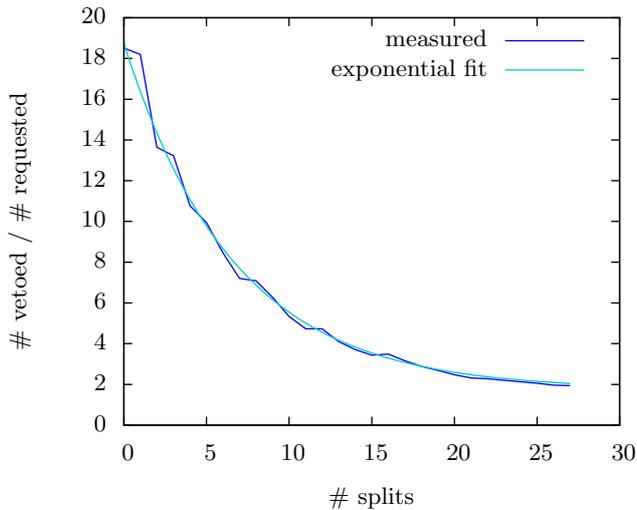}
\caption{\label{figures:performance}Performance of the algorithm
  measured as a ratio of the number of vetoed points to the number of
  events requested as a function of the number of cell splits
  allowed. An exponential improvement is seen as more and more splits
  are considered. In this example, requesting $500000$ events, a
  maximum of $27$ splits occured. The very efficient region from $18$
  splits onward with below three vetoes per generated event has
  been reached after about $40000$ generated events.}
\end{figure}

We also use this example, which closely resembles initial state
backward evolution of a parton shower at small values of the momentum
fraction $x$, to asses the improvements obtained by the adaptive
sampling algorithm. Particularly, we count the number of vetoed points
encountered when requiring the same number of events while limiting
the allowed number of cell splits. This way, a direct comparison of
very coarse to increasingly finer overestimates is performed. The
results are presented in figure~\ref{figures:performance}, showing an
exponential improvement with the number of splits performed.

\section{Conclusions}

The sampling of Sudakov-type distributions is at the heart of all
current parton shower and POWHEG NLO matching implementations.  In
this paper we have introduced the C++ library \program{ExSample},
which targets at adaptive sampling of these distributions as defined
from a splitting kernel, which -- in general -- may only be known
through a function call.

Additional parameters, such as typically encountered dependencies on
incoming parton momentum fractions or the full dependence on a phase
space point governing a hard scattering process, can be dealt with in
full generality. \program{ExSample} has been validated in `toy' as
well as realistic setups, showing full functionality of the
implementation.

\section*{Acknowledgments}
I would like to thank Malin Sj\"odahl and Stefan Gieseke for fruitful
discussion. This work was supported in part by the European Union
Marie Curie Research Training Network MCnet under contract
MRTN-CT-2006-035606 and the Helmholtz Alliance ``Physics at the
Terascale''.

\appendix

\section{Availability and Installation}
\label{sections:install}

\program{ExSample} is available from\\
\vspace*{1ex}

\texttt{http://www.desy.de/\~\ platzer/software/\\\hspace*{6ex}exsample-1.0.tar.gz}\\
\vspace*{1ex}

It is a complete header based library, depending additionally only on
the presence of \program{boost} headers \cite{boost}.  An installation
procedure is thus not required except for making the
\program{ExSample} headers available to the client code during
compilation by including the header file \program{exsample.h}.
\program{ExSample} is published under the GNU General Public License
version 2 \cite{gpl2} and can thus be freely used and redistributed.

The distribution contains extensive documentation, several examples of
usage, as well as an implementation for standard sampling and adaptive
Monte Carlo integration, of which \program{ExSample} is capable as
well.

\bibliography{exsample}

\begin{thebibliography}{10}

\bibitem{Bahr:2008pv}
{M. B\"ahr et al.},
\newblock Eur. Phys. J. {\bf C58}, 639 (2008), 0803.0883.

\bibitem{Sjostrand:2007gs}
T.~Sj{\"o}strand, S.~Mrenna, and P.~Skands,
\newblock Comput. Phys. Commun. {\bf 178}, 852 (2008), 0710.3820.

\bibitem{Gleisberg:2008ta}
T.~Gleisberg {\em et~al.},
\newblock JHEP {\bf 02}, 007 (2009), 0811.4622.

\bibitem{Seymour:1994df}
M.~H. Seymour,
\newblock Comp. Phys. Commun. {\bf 90}, 95 (1995), hep-ph/9410414.

\bibitem{Bengtsson:1986hr}
M.~Bengtsson and T.~Sjostrand,
\newblock Phys. Lett. {\bf B185}, 435 (1987).

\bibitem{Bengtsson:1986et}
M.~Bengtsson and T.~Sjostrand,
\newblock Nucl. Phys. {\bf B289}, 810 (1987).

\bibitem{Norrbin:2000uu}
E.~Norrbin and T.~Sj{\"o}strand,
\newblock Nucl. Phys. {\bf B603}, 297 (2001), hep-ph/0010012.

\bibitem{Miu:1998ju}
G.~Miu and T.~Sjostrand,
\newblock Phys. Lett. {\bf B449}, 313 (1999), hep-ph/9812455.

\bibitem{Nason:2004rx}
P.~Nason,
\newblock JHEP {\bf 11}, 040 (2004), hep-ph/0409146.

\bibitem{Corcella:2002jc}
G.~Corcella {\em et~al.},
\newblock (2002), hep-ph/0210213.

\bibitem{Sjostrand:2006za}
T.~Sj{\"o}strand, S.~Mrenna, and P.~Skands,
\newblock JHEP {\bf 05}, 026 (2006), hep-ph/0603175.

\bibitem{Buckley:2011ms}
A.~Buckley {\em et~al.},
\newblock Phys. Rept. {\bf 504}, 145 (2011), 1101.2599.

\bibitem{Platzer:2011dq}
S.~Platzer and M.~Sjodahl,
\newblock Eur.Phys.J.Plus {\bf 127}, 26 (2012), 1108.6180.

\bibitem{Lonnblad:ACDC}
L.~L{\"o}nnblad,
\newblock {ACDC -- The Auto Compensating Divide-and-Conquer Phase Space
  Generator}.

\bibitem{Jadach:2002kn}
S.~Jadach,
\newblock Comput. Phys. Commun. {\bf 152}, 55 (2003), physics/0203033.

\bibitem{boost}
\texttt{http://www.boost.org}.

\bibitem{gpl2}
\texttt{http://www.gnu.org/licenses/}.

\end{thebibliography}

\end{document}